\documentclass[a4paper,11pt]{article}
\usepackage{jinstpub} 
\usepackage{lineno}
\usepackage{caption}
\usepackage{subcaption}
\usepackage{comment}
\graphicspath{{figures/}}
\usepackage{xcolor}
\usepackage{soul}
\usepackage{svg}
\usepackage{hyperref}

\title{\boldmath Development and characterization of monolithic sensors with internal gain in a 110~nm CMOS process}

\author[a,c,1]{G. Gioachin\note[1]{\label{author}Corresponding authors.},}
\author[a,1]{C. Ferrero,}
\author[d]{M. Bregant,}
\author[a,c]{S. Bufalino,}
\author[m]{Z. Buthelezi,}
\author[f]{F. Carnesecchi,}
\author[f]{D. Cavazza,}
\author[a,e]{M. Ciacco,}
\author[f,g]{M. Colocci,}
\author[a]{M. Da Rocha Rolo,}
\author[l]{G. Di Nicolantonio,}
\author[a]{S. Durando,}
\author[h,i,p]{J. Endrizzi,}
\author[f,g]{G. Fabbri,}
\author[m]{S. Foertsch,}
\author[a,b]{U. Follo,}
\author[n]{J. Goodhead,}
\author[f,g]{M. Grenno,}
\author[p]{H. Iizuka,}
\author[p]{B. Lim,}
\author[a]{M. Mandurrino,}
\author[l]{G. Margutti,}
\author[a,e]{M. Masera,}
\author[a]{M. Mignone,}
\author[d]{M. G. Munhoz,}
\author[f]{R. Nania,}
\author[m]{F. G. Ntshangase,}
\author[a,b]{V. Pagliarino,}
\author[h,i]{L. Pancheri,}
\author[a,e]{F. Picollo,}
\author[a]{A. Rivetti,}
\author[f,g]{B. Sabiu,}
\author[f,g]{G. Scioli,}
\author[d,2]{G. G. A. de Souza\note[2]{Now at Brazilian Nanotechnology National Laboratory (LNNano), Brazilian Center for Research in Energy and Material (CNPEM), Campinas 13083-970, Brazil},}
\author[f,g]{S. Strazzi,}
\author[f,g]{L. Tomba,}
\author[n]{S. L. Winberg,}
\author[a]{R. Wheadon}

\emailAdd{giulia.gioachin@to.infn.it}
\emailAdd{chiara.ferrero@to.infn.it}

\affiliation[a]{Istituto Nazionale di Fisica Nucleare (INFN), Sezione di Torino, Via P. Giuria 1, 10125, Torino, Italy}
\affiliation[b]{Dipartimento di Elettronica e Telecomunicazioni (DET), Politecnico di Torino, Corso Duca degli Abruzzi 24, 10129, Torino, Italy}
\affiliation[c]{Dipartimento di Scienza Applicata e Tecnologia (DISAT), Politecnico di Torino, Corso Duca degli Abruzzi 24, 10129, Torino, Italy}
\affiliation[d]{Instituto de F\'{i}sica da Universidade de S\~{a}o Paulo, Rua do Mat\~{a}o 1371, 05508-090 Cidade Universit\'{a}ria, S\~{a}o Paulo, Brazil}
\affiliation[e]{Dipartimento di Fisica, Università di Torino,  Via Pietro Giuria, 1, Torino, 10125, Italy}
\affiliation[f]{Istituto Nazionale di Fisica Nucleare (INFN), Sezione di Bologna, Viale C. Berti Pichat 6/2, 40127, Bologna, Italy}
\affiliation[g]{Dipartimento Fisica e Astronomia, Università di Bologna, Viale C. Berti Pichat 6/2, 40127, Bologna, Italy}
\affiliation[h]{Dipartimento di Ingegneria Industriale, Universit\`{a} di Trento, Via Sommarive 9, 38123, Trento, Italy}
\affiliation[i]{TIFPA-INFN, Via Sommarive 14, 38123, Trento, Italy}
\affiliation[l]{LFoundry S.r.l, Via Pacinotti 7, 67051, Avezzano (AQ), Italy}
\affiliation[]{Brazilian Nanotechnology National Laboratory (LNNano), Brazilian Center for Research in Energy and Material (CNPEM), Campinas 13083-970, Brazil}
\affiliation[m]{iThemba LABS, National Research Foundation, Somerset West, South Africa}
\affiliation[n]{University of Cape Town: Cape Town, Western Cape, South Africa}
\affiliation[o]{Fondazione Bruno Kessler, Via Sommarive 18, 38123, Trento, Italy}
\affiliation[p]{Center for Nuclear Study, The University of Tokyo, Tokyo, 113-0033, Japan}

\abstract{
This paper presents the optimization and characterization of a Monolithic Active Pixel Sensor with internal gain implemented in the 110~nm CMOS process from LFoundry. 
This work, carried out within the framework of the latest ARCADIA production run, reports on the characterization of sensors with an optimized gain layer. Improvements over earlier prototypes, including substrate current reduction via guard-ring connection, are briefly discussed. Both laboratory and in-beam measurements are presented.
Using soft X-rays from a $^{55}$Fe source, both the pixel response and the noise performance are analyzed. In addition, laser measurements demonstrate an electronic jitter ranging from 20 to 40~ps for a signal amplitude corresponding to one minimum ionizing particle (MIP). Finally, the results from several test beam campaigns are described, showing the signal amplitude, gain and time resolution of sensors with different doping profiles of the gain layer evaluated with MIPs. A time resolution of about 78~ps was measured in the optimal condition of bias voltage and electronics configuration. The results demonstrate that monolithic CMOS sensors with internal gain can achieve promising timing performance at ambient temperature, highlighting their strong potential for next-generation precision timing applications.
}

\keywords{Solid-state detectors, Timing detectors, Front-end electronics for detector readout, Manufacturing}

\newpage
\begin{document}
\maketitle
\flushbottom

\section{Introduction}
\label{sec:intro}
Monolithic Active Pixel Sensors (MAPS), implemented using standard CMOS processes, are currently one of the most promising technologies for a wide range of applications, including high-energy physics, medical imaging, and space experiments~\cite{AGLIERIRINELLA2023_DPTS,AGLIERIRINELLA2024_APTS,Limadou}. 
MAPS offer several advantages over traditional hybrid pixel detector technologies. In particular, they can be fabricated at relatively low cost even for large-area arrays, while maintaining a low material budget and a reduced power consumption. 
The upcoming high-energy physics experiments and future particle detectors will demand more stringent requirements, including handling unprecedented pile-up levels~\cite{HL-LHC, FCC}. Consequently, detectors will be required to provide enhanced spatial granularity and excellent time resolution in order to efficiently disentangle overlapping interactions and ensure efficient event reconstruction and background discrimination. Moreover, an outstanding time resolution is crucial for Particle IDentification (PID), where precise timing information will significantly improve the separation of particle species over a broad momentum range. In this context, current developments aim at achieving time resolutions of about 30-40~ps for experiments such as ATLAS and CMS~\cite{CMS-MTD-TDR,ATLAS-HGTD-TDR}, and down to approximately 20~ps for next-generation detectors like ALICE 3~\cite{LoI_ALICE3}\cite{scoping_doc}. 
Overall, these challenges necessitate significant technological advancements in detector design, and the scientific community is actively working to enhance the timing performance.\\
Currently, monolithic technologies are the primary choice for tracking detector applications, where outstanding efficiency, spatial resolution, low material budget, and high radiation hardness are key requirements. For instance, the ALPIDE sensors employed in the ALICE ITS2 have demonstrated a detection efficiency above 99\% after irradiation up to a non-ionizing fluence of \(1.7~\times~10^{13}\)~1~MeV~\(n_{\mathrm{eq}}/\mathrm{cm}^2\)~\cite{alpide}. More recently, Analog Pixel Test Structures (APTS) developed in 65~nm CMOS technology for the ALICE ITS3 upgrade have maintained detection efficiencies above 99\% up to a fluence of \(10^{14}\) 1~MeV \(n_{\mathrm{eq}}/\mathrm{cm}^2\), one order of magnitude above the ITS3 radiation hardness requirement~\cite{AGLIERIRINELLA2024_APTS}. 
However, no large-scale monolithic demonstrator has been produced for HEP applications with sub 50~ps timing resolution requirements, such as the case of the Time-Of-Flight (TOF) detector of the proposed ALICE 3 experiment. Several R\&D projects are currently underway to demonstrate the feasibility of achieving timing resolutions of O(10~ps) with monolithic devices~\cite{APTS-OA,Paolozzi_2025, minicactus65pstime}. Nevertheless, this timing performance is currently achieved only by prototype implementations, which do not yet fulfill the requirements of large-scale experiments, such as low power consumption and scalability to large areas. Indeed, in some cases, the power consumption remains prohibitively high, or the front-end electronics is integrated outside the pixel to reduce the sensor capacitance.\\
The three main ingredients to achieve good timing performance are: 
\begin{itemize}
    \item a fast charge collection, enabled by a full depletion of the sensor;
    \item  the optimization of the device geometry and the electric field profile to reduce the distortion term contribution to the time resolution~\cite{Riegler_2017};
    \item a high Signal-to-Noise Ratio (SNR), which can be increased by adding a gain layer below the collection electrode, enabling charge multiplication, as in the case of the Low Gain Avalanche Diode (LGAD).
\end{itemize}

\noindent The integration of the LGAD concept~\cite{Sadrozinski_2018} in a monolithic technology was already proven in the third ARCADIA (Advanced Readout CMOS Architectures with Depleted Integrated Sensor Arrays) Engineering Run (ER3)\footnote{An engineering run consists of a dedicated fabrication cycle specifically designed for process development and optimization, rather than for large-scale manufacturing of final devices. In this context, the production is intentionally configured to probe and validate specific process variations (among them, the doping profiles), aiming at refining the technology prior to full-scale deployment. In the case of CMOS-LGAD technologies, the fabrication flow requires a complete set of masks for the fabrication of the CMOS transistors, complemented by an additional dedicated mask for the LGAD gain layer implant, as well as optimized wafer substrates. Since the majority of the masks are associated with transistor fabrication, existing mask sets, such as those developed within the ARCADIA project~\cite{ARCADIA}, can be reused, significantly reducing production costs.\\}, where a gain was introduced in fully depleted MAPS produced in the commercial 110~nm CMOS Image Sensor (CIS) process by LFoundry~\cite{Follo_2024, Pancheri_2020}.
\noindent The first production in ER3 showed a sensor gain of approximately 2.5, which was measured both for passive structures without integrated electronics~\cite{2024Corradino_JINST} and in a sensor prototype with internal gain and in-pixel front end electronics: the Monolithic CMOS Avalanche Detector Pixelated (MADPix)~\cite{Follo_2024}. Subsequently, a new Short-Loop (SL) version was submitted to the foundry with an optimized gain layer dose. In this SL, the same mask set of the ER3 was used, enabling both an optimization of the sensor and an evaluation of the impact of some process variations on the device performance.\\

In this paper, the results of the characterization of the new ARCADIA short-loop are presented, illustrating how the critical issues of the first production were addressed. 
In particular, Section~\ref{FIB} describes how the substrate current of the ER3 monolithic sensors was reduced by performing an electrical connection with a Focused Ion Beam (FIB) technique, which was later also applied to the structures of the new ARCADIA SL. In Section~\ref{run3.2}, a new SL production with an enhanced gain value is presented, aiming to further improve the timing performance. In Section~\ref{Laboratory measurements}, the results of sensor characterization in the laboratory are illustrated. Section~\ref{Beam test} presents the latest test beam measurements carried out at the Proton Synchrotron (PS) at CERN, along with the results obtained from the analysis of the collected data. Finally, the future perspectives and planned advancements of the technology are outlined in Section~\ref{Conclusions}.

\section{Detector under test}

\begin{figure}[!ht]
    \centerline{\includegraphics[width=0.75\linewidth]{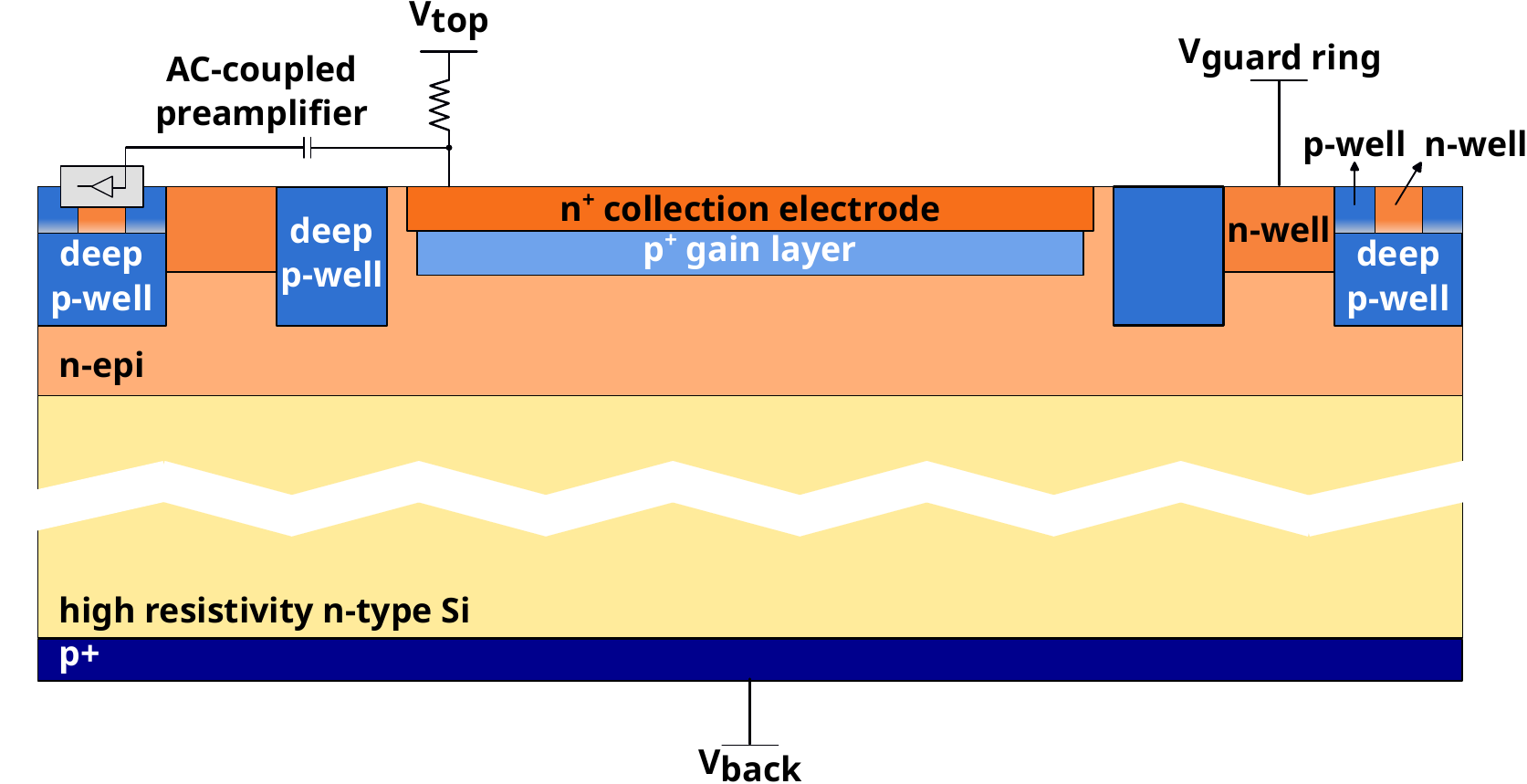}}
    \caption{Schematic cross-section of one pixel, not to scale, of the monolithic sensor with the added gain layer produced by the ARCADIA collaboration.}
    \label{fig:cross-section}
\end{figure}

The innovative concept of the MADPix prototype, described in detail in \cite{Follo_2024}, was implemented in a monolithic sensor composed of eight matrices, each consisting of 64 pixels with a size of 100~$\times$~250~\textmu m$^2$. Figure~\ref{fig:cross-section} illustrates the ARCADIA sensor concept with the addition of a gain layer. 
The device is based on a shallow n$^+$ collection electrode, and a p$^+$ gain layer, just beneath the pad, with the purpose to generate a localised high electric field (with a peak in the order of 300-400~kV/cm), enabling to trigger charge multiplication by impact ionization. The n-type guard-ring structures are implemented to prevent charge carriers, generated by particles impinging far from the active region, from reaching the gain layer. This mitigation is essential to avoid a slowing down of the signal formation and a consequent degradation of the timing performance. The front-end electronics is embedded in deep p-wells surrounding the collection region, ensuring isolation and proper charge collection. The active volume consists of a double n-type epitaxial layer with a total thickness of about 48~\textmu m, grown on a highly doped p$^+$ substrate that enables back-side biasing and completes the junction.\\

\noindent Four bias voltages must be supplied to the sensor to set its working point.
A negative bias (typically between -25~V and -30~V) is applied to the backside p$^{+}$ implant ($V_{\mathrm{back}}$) to fully deplete the active region and enable charge collection by drift. Moreover, a positive voltage (in the range between 50~V and 70~V) is applied to the topside electrode ($V_{\mathrm{top}}$) to control the gain. By depleting the gain implant region, the electric field in the gain layer becomes high enough to trigger the onset of the charge multiplication. The higher the voltage applied to the collection electrode, the higher the gain. The guard-rings are biased at 8–9~V to ensure effective shielding and, in combination with the backside bias, to achieve full depletion of the substrate. Finally, the deep p-wells are kept at 0~V, thereby ensuring also the common ground for the other contacts.

\begin{figure}[ht]
    \centerline{\includegraphics[width=0.8\linewidth]{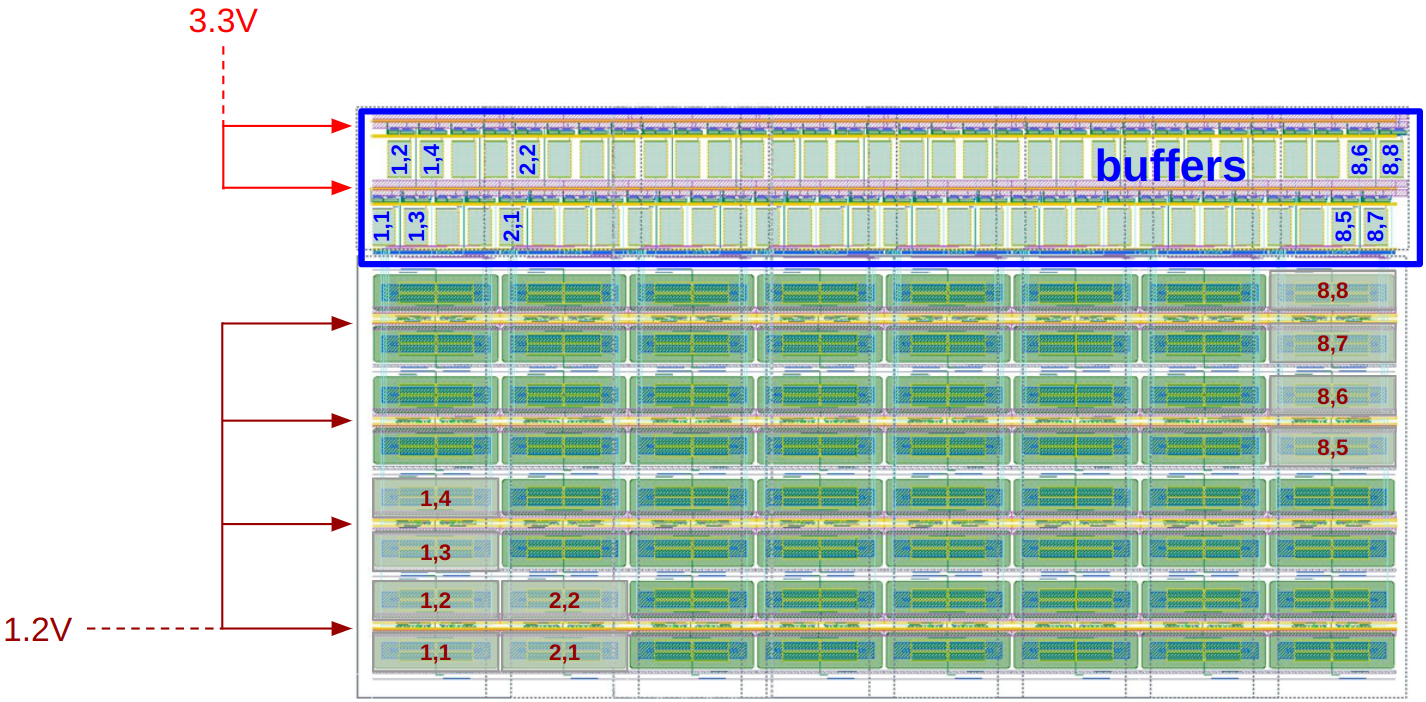}}
    \caption{A MADPix matrix layout with the array of 8~×~8 pixels and the readout buffers powered at 3.3~V outside the matrix (highlighted in blue). The four yellow horizontal stripes represent the 1.2~V power rails of the front-end electronics. }
    \label{fig:MADPix_layout}
\end{figure}

\noindent The electronics of the chip is implemented in a 110~nm CMOS technology node and the amplifier design is based on a cascoded common source architecture.
Each in-pixel front-end is then linked to a buffer, used to drive and read out the pixel signal, implemented as a source follower placed outside the matrix, as can be seen in Fig.~\ref{fig:MADPix_layout}. This figure shows a matrix of 64 pixels organized in an 8~×~8 array. Each pixel is identified by its column–row coordinates, corresponding to its position within the array.
The in-pixel front-end operates at a 1.2~V, which is introduced from the bottom side of the matrix and distributed to all pixels via four horizontal metal lines (highlighted in yellow inside the matrix). This results in a left-to-right power distribution across the array. On the other hand, the peripheral buffers operate at 3.3~V. In this case, the supply is provided from the topside of the matrix and it is split into two lines to power the buffers. Each pixel is associated with a dedicated buffer. This one-to-one correspondence is highlighted by the coordinate-based labelling, which is consistently replicated in the buffer array, as shown in Fig.~\ref{fig:MADPix_layout}.

\noindent The power consumption of the electronics is reported henceforth at the nominal bias of the front-end (standard configuration) and at +20\% of the voltage domain, with the supply voltage increased from 1.2~V to 1.5~V and the front-end current of $\approx +25\%$. In the first case, a power consumption of approximately 0.18~mW per channel was measured, corresponding to about 6.7~mW/mm$^{2}$. In the second front-end configuration, the power consumption increases to approximately 0.28~mW per channel, equivalent to about 11~mW/mm$^{2}$. Nevertheless, since the front-end was originally designed for a final pixel area of 250~\textmu m~$\times$~250~\textmu m, the power consumption would be lower in the standard configuration for this larger pixel size.

\subsection{Mitigation of the substrate current}
\label{FIB}

\begin{figure}[!ht]
    \centering
    \begin{subfigure}[!ht]{0.49\linewidth}
        \centering
        \includegraphics[width=\textwidth]{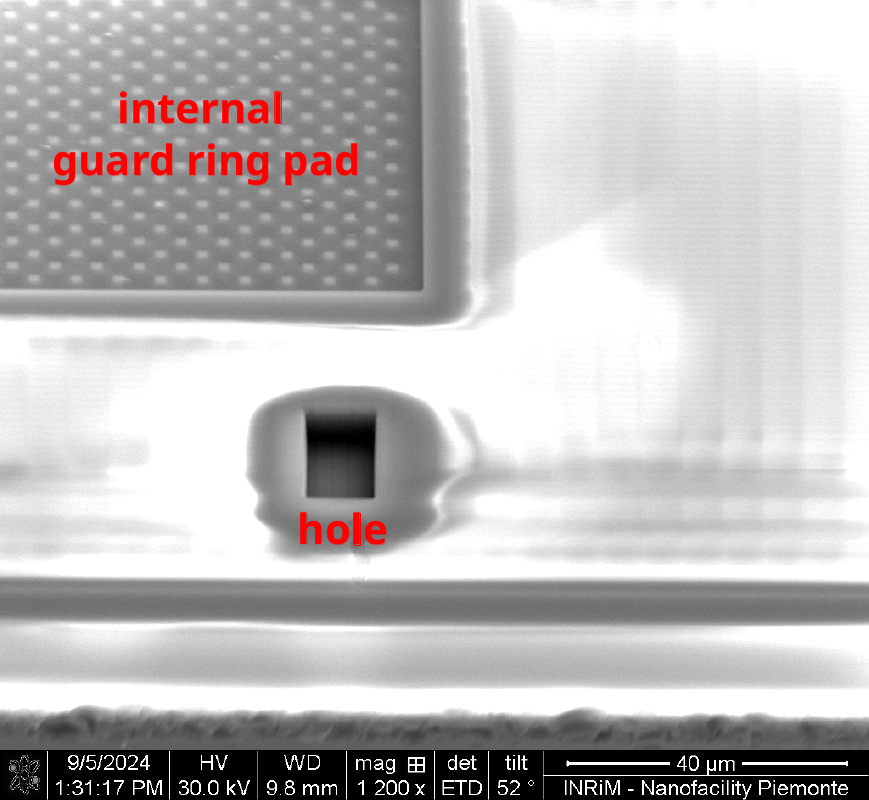}
        \caption{}
        \label{fig:hole}
    \end{subfigure}
    \begin{subfigure}[!ht]{0.49\linewidth}
        \centering
        \includegraphics[width=\textwidth]{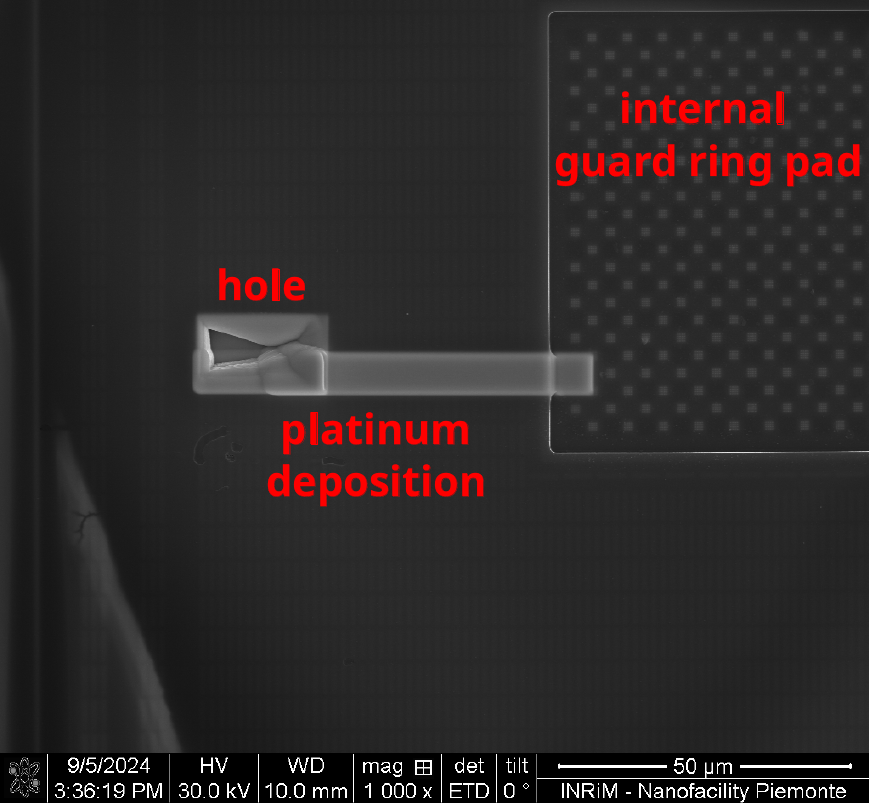}
        \caption{}
        \label{fig:platinum}
    \end{subfigure}
    \caption{Electron microscope images of (a) the hole created with the ion beam (centre of the image) and (b) the platinum deposition which is approximately 1~\textmu m thick and 30~\textmu m long (light gray). In (b), the pad devoted to bias the internal guard-ring is shown on the right. The figure is rotated by 90° with respect to (a).}
    \label{fig:layout}
\end{figure}

As mentioned in~\cite{Follo_2024}, a high substrate current was measured in the monolithic prototype of the ER3 compared to the passive structures~\cite{2024Corradino_JINST}.
This unexpected behavior of the prototype was attributed to an unconnected guard-ring placed between the pad frame and the dicing perimeter, which was supposed to be biased at a positive voltage below 10~V to avoid border effects.
Employing a setup located at the Istituto Nazionale di Ricerca Metrologica (INRiM) in Torino, the floating guard-ring was connected to the proper pad through the Focused Ion Beam (FIB) technique~\cite{FIB}. During the FIB process, the silicon dioxide layer protecting the chip surface was locally removed using a gallium ion beam, exposing the guard-ring metal line (Fig.~\ref{fig:hole}). Subsequently, a platinum layer was deposited to electrically connect the metal line of the chip to the pad used to bias the internal guard-ring, as shown in Fig.~\ref{fig:platinum}. Since the current drawn by the external guard-ring is $<0.01$~mA, the IR drop due to the metal line is negligible even if the connection is thin.

\begin{figure}[ht]
    \centerline{\includegraphics[width=0.5\linewidth]{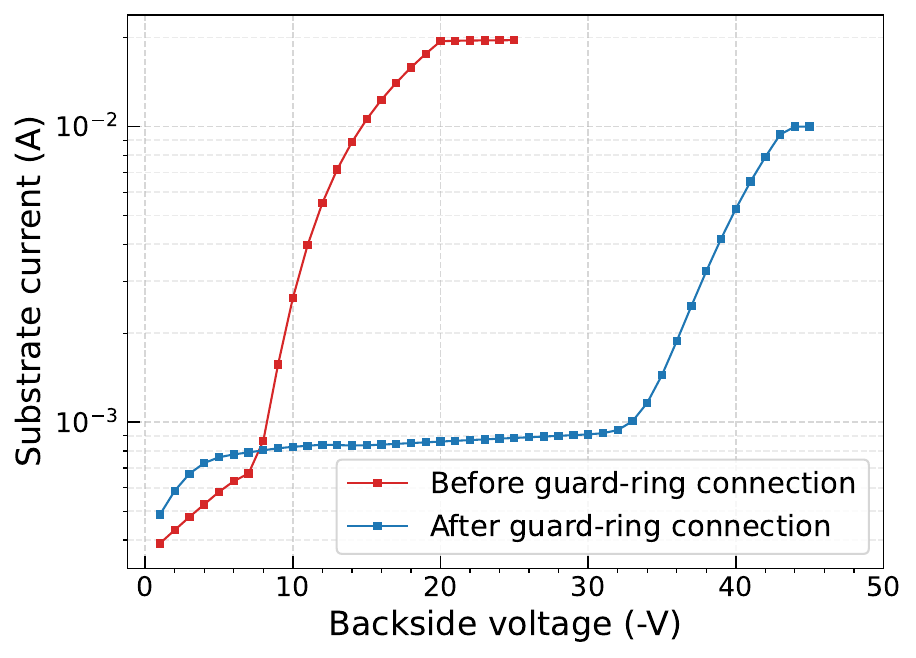}}
    \caption{Substrate current versus backside voltage before and after the guard-ring connection for the ER3 sensor. The top collection electrode was biased at 45~V while the guard-ring was at 8.5~V. After the guard-ring connection, the punch-through effect starts above -35~V.}
    \label{fig:I-V_before_after_FIB}
\end{figure}

\noindent The samples were subsequently wire-bonded to the test boards, and Current versus Voltage (I-V) measurements were performed to evaluate whether the excessive substrate current had been mitigated after FIB. The results are shown in Fig.~\ref{fig:I-V_before_after_FIB}. For backside voltages above 15-20~V the sample with the connected guard-ring exhibits a current approximately two orders of magnitude lower than that of the chip with floating guard-ring. This behavior is consistent with measurements on passive structures~\cite{2024Corradino_JINST}, where the external guard-ring was connected to the internal one.\\

\noindent The same guard-ring connection procedure was consistently applied to the structures of the new ARCADIA short-loop, which will be described in the following section.

\subsection{Gain layer optimization in the new short-loop production}
\label{run3.2}

In order to achieve an optimal time resolution, an internal gain around 20 is required, as demonstrated in~\cite{Sadrozinski_2018}. To meet this target, the multiplication layer profile was optimized based on the first production, and a short-loop was submitted. Specifically, the ion implantation energy was increased by approximately 30\% compared to the previous run, resulting in a deeper gain implant designed to achieve the required gain.\\
Sensors from the new SL were fabricated in 2024, reusing the same engineering mask set adopted for the previous ER3. The doping concentration of the gain implant was increased to compensate for the reduced gain observed in the first production~\cite{Follo_2024}. Synopsys Sentaurus Technology CAD (TCAD)~\cite{Synopsys_TCAD} simulations played a key role in this phase, as they were used to identify the appropriate gain doses required to achieve the target gain.
In order to account for possible process variations, wafers with different dose splits were produced: a reference split (hereafter called +0\%) and five additional splits with a dose of -10\%, -6.7\%, -3.3\%, +3.3\% and +6.7\% with respect to the reference one. In the following, only the results obtained for the -3.3\%, 0\%, and +3.3\% gain layer dose splits are reported, as these allow the desired gain, ranging from 6 to 20 with TCAD simulations, to be achieved.

\section{Laboratory characterization results} 
\label{Laboratory measurements}
The new devices were characterized in the laboratory at ambient temperature through electrical, $^{55}$Fe, and laser measurements.
As explained in~\cite{Follo_2024}, only 4 pixels out of 64 of a matrix can be read out at the same time by a custom PCB Test Board. The board features four analog buffers driving 50~$\Omega$ lines acquired through an oscilloscope via SMA cables. 

\subsection{Current-Voltage measurements}

\begin{figure}[!ht]
    \centering
    \begin{subfigure}[!ht]{0.49\linewidth}
        \centering
        \includegraphics[width=\textwidth]{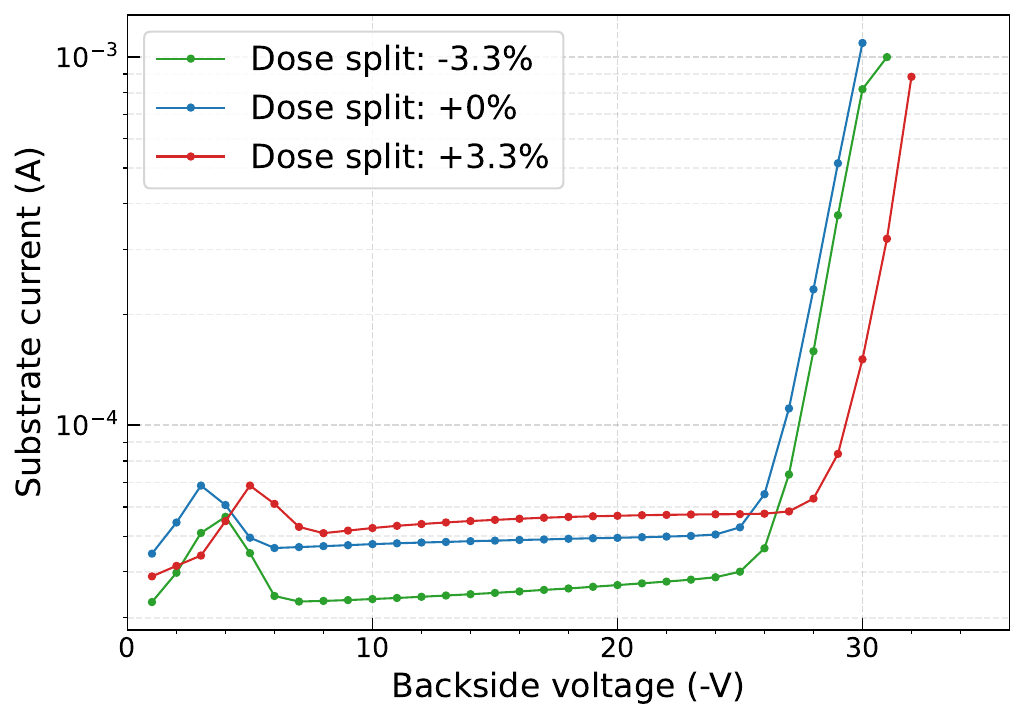}
        \caption{}
        \label{fig:back_back}
    \end{subfigure}
    \begin{subfigure}[!ht]{0.49\linewidth}
        \centering
        \includegraphics[width=\textwidth]{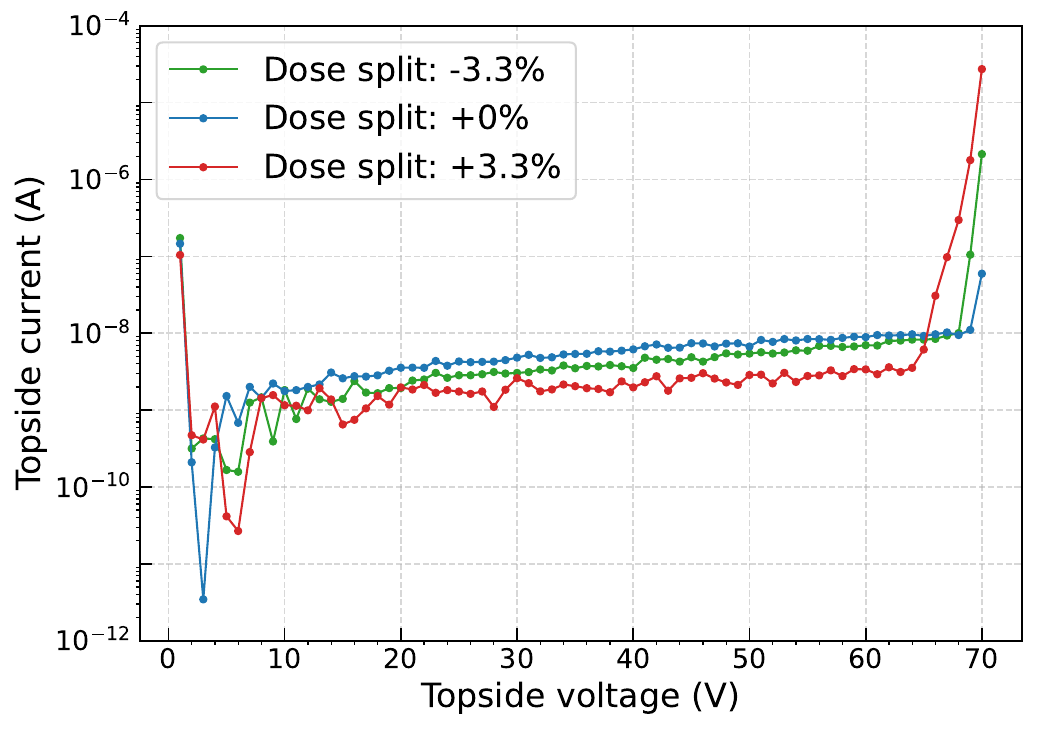}
        \caption{}
        \label{fig:top_top}
    \end{subfigure}
    \caption{(a) The substrate current as a function of backside voltage for three gain-dose splits. The collection electrode voltage was biased at 45~V, while the guard-ring was at 8.5~V. (b) The topside current as a function of the topside voltage. The backside electrode was biased at -25~V while the guard-ring was at 8.5~V.}
    \label{fig:IV_run3.2}
\end{figure}

The substrate (Fig.~\ref{fig:back_back}) and topside electrode currents (Fig.~\ref{fig:top_top}) were measured as a function of the backside and topside voltage, respectively, for three different gain layer dose splits: -3.3\%, +0\%, +3.3\%.\\
\noindent Figure~\ref{fig:back_back} illustrates the onset of punch-through phenomenon in the range of approximately -25~V to -28~V, where the current begins to increase exponentially due to a hole current flowing between the frontside p implants and the p$^+$ doped electrode located at the bottom of the sensor. The minimum current measured in the plateau region is lower than that observed in Fig.~\ref{fig:I-V_before_after_FIB} after the FIB process. This current difference arises from the different dicing methods used in the two cases: the devices from the short-loop were laser diced, whereas in the ER3 the blade dicing was used. These two dicing techniques affect the sensor edges differently: specifically, sawing tends to introduce more defects, which in turn lead to higher substrate leakage currents.\\
Figure~\ref{fig:top_top} shows the current measured at the topside collection electrode as a function of its bias voltage. 
The breakdown effect is observed in the voltage range of approximately 65–70~V, where the device enters a regime of uncontrolled carrier multiplication, resulting in a significant increase in leakage current. As expected, the breakdown voltage for the +3.3\% gain dose split is reached at a lower voltage due to the higher gain.

\subsection{Pixel uniformity using $^{55}$Fe source}
\label{sec:Fe}
\begin{figure}[!ht]
    \centering
    \begin{subfigure}[!ht]{0.49\linewidth}
        \centering
        \includegraphics[width=\textwidth]{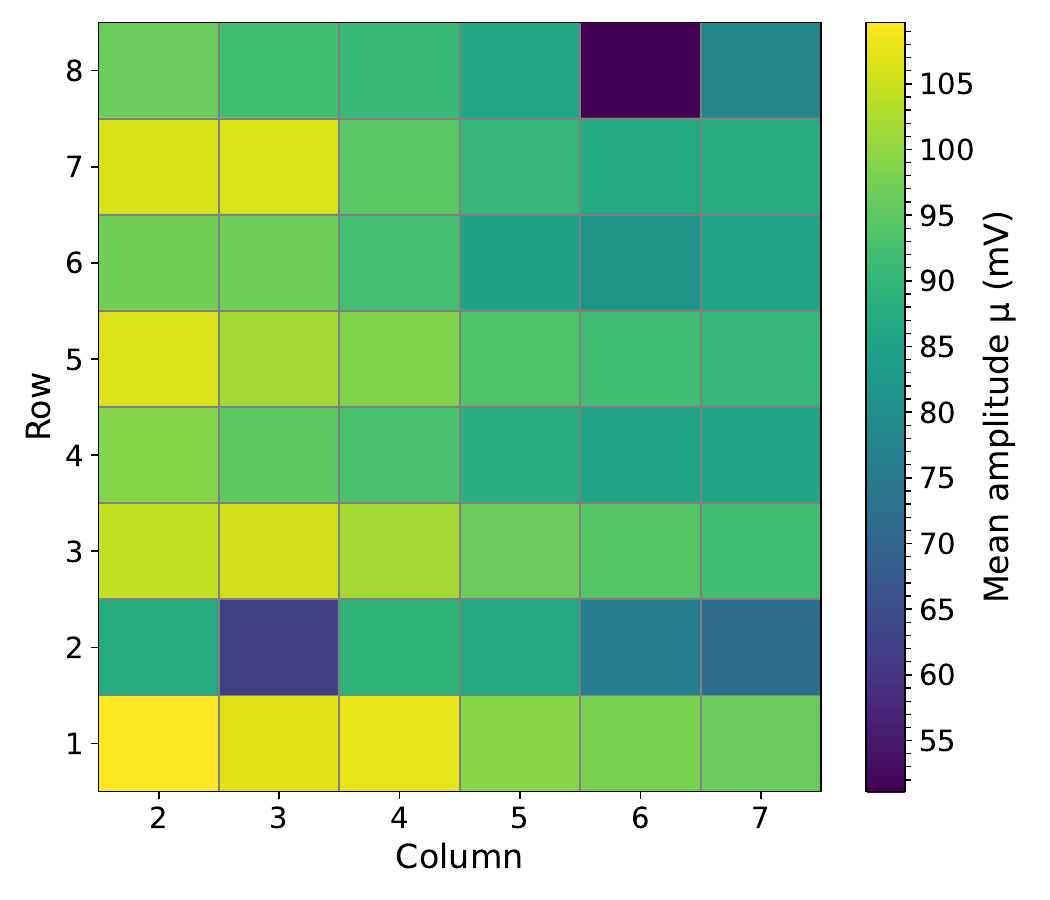}
        \caption{}
        \label{fig:amp_55Fe}
    \end{subfigure}
    \begin{subfigure}[!ht]{0.49\linewidth}
        \centering
        \includegraphics[width=\textwidth]{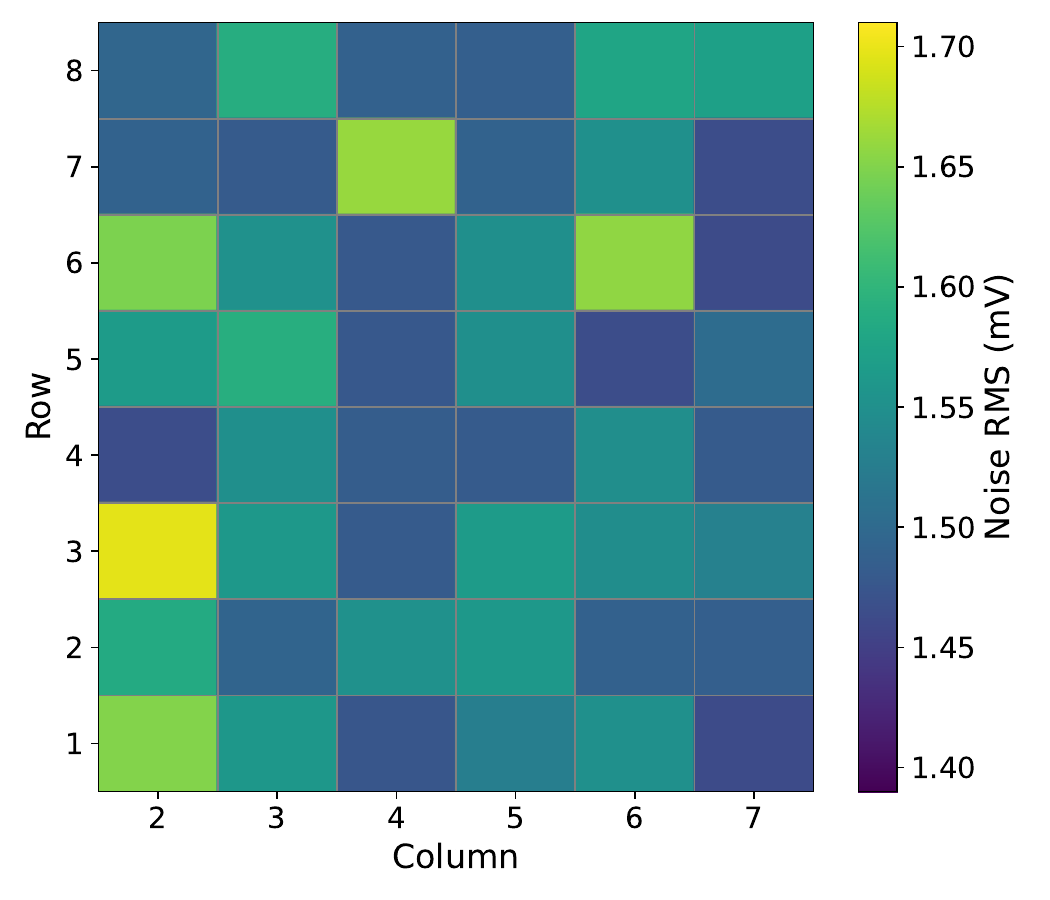}
        \caption{}
        \label{fig:noise_55Fe}
    \end{subfigure}
    \caption{(a) Map of the mean amplitude values for different pixels, employing X-rays from a $^{55}$Fe source. (b) Map of the RMS of the noise distribution of each pixel. The measurements were performed biasing the backside electrode at -25~V, the guard-ring at 9~V, the topside voltage at 60~V and in the standard front-end electronics configuration.
    }
    \label{fig:55Fe}
\end{figure}

A sensor with gain dose split +3.3\% was further characterized in the laboratory by investigating its response to a non-collimated $^{55}$Fe monochromatic soft X-ray source. The source~\cite{Fe_radiation_sources} emits photons mainly with an energy of 5.9~keV~\cite{55Fe} and it features an active diameter of 10~mm with 37~MBq nominal activity.\\
The goal of this study was to evaluate the response uniformity of pixels within one chip matrix by measuring the signal amplitudes induced by $^{55}$Fe photons. 
This analysis is used to assess the impact of the IR voltage drop of the electronics power along the matrix, which is generally not negligible and may lead to non-uniform signal amplitudes across the chip. Subsequently, the noise was also evaluated on a pixel-by-pixel basis to further characterize the uniformity of the detector response.
The data acquisition was performed by employing a Tektronix Low Profile Digitizer~\cite{tektronix_6series_digitizer} featuring a sampling rate of 25~GS/s and 8~GHz bandwidth.\\
Figure~\ref{fig:amp_55Fe} shows the map of the mean signal amplitude measured for 48 pixels within the matrix, excluding the peripheral columns, for which measurements were not performed. For each pixel, the signal amplitude distribution is fitted with a Gaussian function in an asymmetric interval around the peak to suppress the influence of the tails. The resulting mean, $\mu$, represents the characteristic amplitude of that pixel. The observed variations in signal amplitude across the matrix can be attributed to different sources depending on the spatial direction. 

\noindent Variations along the vertical direction are likely caused by the higher resistance (capacitance) of the longer routing paths required to reach the more distant buffers located farther from the matrix (see the upper buffer row in Fig.~\ref{fig:MADPix_layout}), which attenuate the signal and result in smaller amplitudes. Since the buffers are arranged in two alternating rows, this effect leads to the observed even/odd row amplitude pattern. In contrast, variations along the horizontal direction can be explained by IR voltage drops along the electronics power rails, visible in Fig.~\ref{fig:MADPix_layout}. Quantitatively, the standard deviation of the amplitude across all the pixels is approximately 10~mV, corresponding to about 12\% of the mean value. If, instead, the maximum semidispersion of 30~mV is considered, this represents roughly 30\% of the mean. Figure~\ref{fig:noise_55Fe} illustrates the RMS of the noise distribution across the matrix, computed on a single point located before the signal region. The noise map shows a good uniformity across the pixel matrix, with a mean noise of 1.53~mV and a pixel-to-pixel standard deviation of 0.06~mV. Individual values range between 1.46 and 1.70~mV, and the statistical uncertainty on each pixel is below 0.01~mV.

In conclusion, the observed horizontal and vertical amplitude differences provide valuable insight into the electrical behavior of the system. These significant findings offer quantitative information that can guide design optimizations and enable corrections in future ERs.

\subsection{Laser characterization}
\label{sec:laser}
The behavior of sensors with a gain dose split of +3.3\% was studied using a picosecond pulsed laser with a wavelength of 1054~nm (PiLas PiL036XSM). These measurements directly allowed to extract the contribution of the electronic jitter to the overall time resolution. As discussed in~\cite{Follo_2024}, in the timing measurements with laser, the dominant contribution arises from the electronic jitter, while the Landau and Distortion terms can be considered negligible for photons and a small laser spot with respect to the pixel size.\\
The analog signal from one pixel was acquired with a LeCroy WaveRunner 9404M-MS oscilloscope~\cite{lecroy_waverunner_9404m}, with a 20~Gs/s sampling rate and an analogue bandwidth of 4~GHz. The laser beam, which was delivered through a single-mode optical fiber, was collimated and subsequently focused onto a micrometric spot of about 8~\textmu m diameter. The device alignment was performed employing a two-axis micrometric translation stages.\\
\noindent To estimate the electronic jitter contribution, the Leading-Edge (LE) algorithm was employed. In this approach, the photon arrival time is defined as the instant when the signal crosses a fixed threshold. The time-walk effect, due to a non-uniform laser response, is corrected a posteriori through an amplitude-based correction derived from the Device Under Test (DUT) signal. The function employed to correct the effect is the same as described in detail in~\cite{Follo_2024}, where each event was analyzed by computing the time difference between the photon arrival time and a reference time $t_0$ provided by the laser source. The intrinsic jitter of the laser, about 4~ps, was considered negligible. Finally, the time resolution $\sigma_{t}$ was defined as the standard deviation obtained from a Gaussian fit on the time difference distributions. The threshold was optimized by selecting the value that minimizes the measured time resolution.\\ 

\begin{figure}[!ht]
  \centering
  \includegraphics[width=0.5\textwidth]{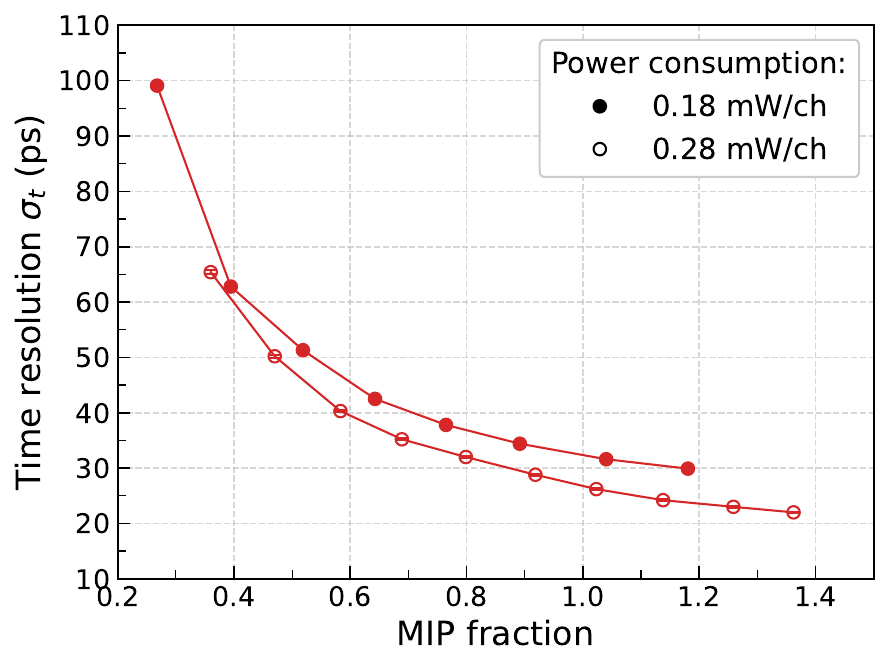}
  \caption{Time resolution from the experimental laser measurement as a function of the signal amplitude normalized to 1 MIP, in the standard configuration (0.18~mW/ch) and in the configuration with increased electronics supply voltage (0.28~mW/ch). The MIP amplitude is about 200~mV in the first case and 230~mV in the latter one. Measurements were taken by fixing the backside electrode at -25~V, the topside voltage at 66~V, and the guard-ring at 9~V. The associated statistical uncertainty is not visible, but it corresponds to about 0.5\% of the $\sigma_{t}$ value.}
  \label{fig:jitter}
\end{figure}

\noindent To evaluate the time resolution through laser measurements, several amplitudes were obtained by varying the attenuation of the laser light. 
Subsequently, they were normalized to the MIP amplitude, which was measured during a test beam campaign and found to be approximately 200~mV in the standard configuration. This value increased to 230~mV when the power consumption was raised to 0.28~mW/ch. Therefore, Fig.~\ref{fig:jitter} shows the time resolution as a function of the signal amplitude normalized to the MIP value for the standard configuration (0.18~mW/ch) and the configuration with increased electronics supply voltage. The jitter estimated for a MIP signal is below 35~ps in the standard configuration, and it improves to $\sim$25~ps in the higher amplification configuration.

\section{Measurements with charged-particle beams} 
\label{Beam test}
\subsection{Test beam setup and analysis method}
Beam measurements were carried out at the CERN Proton Synchrotron (PS) in the T10 beam line at ambient temperature. The experimental setup is described in detail in~\cite{Follo_2024}. The beam was mainly composed of pions and protons with momentum of 10~GeV/$c$. 
The time reference was provided by an LGAD sensor, previously characterized in test beam campaigns~\cite{carnesecchi2023_doubleLGAD, carnesecchi2022beamtestresults25}, which was installed in the upstream telescope plane position. This sensor was bonded on a dedicated amplifier board (Single-channel SC board~\cite{2021Ferrero_CRC}), followed by a high-bandwidth amplifier (Minicircuit LEE39+~\cite{MiniCircuits_LEE-39+}).
The DUT, MADPix, was positioned in the subsequent plane. The analog signals from three DUT pixels were acquired with a LeCroy WaveRunner 9404M-MS oscilloscope. The remaining channel of the oscilloscope was devoted to read out the LGAD.
The trigger was defined as the logical OR of the three DUT pixels, with the threshold properly chosen in order to suppress bipolar signals generated by particles impinging on neighbouring pixels.\\
The collected data were analyzed offline by exploiting the full waveform information. A digital filter based on a 4-point moving average was applied to reduce high-frequency noise.\\
To ensure a clean data sample, waveforms were discarded if at least one of the following conditions was met:
\begin{itemize}
    \item the maximum amplitude in a time window of O(10~ns) prior to the expected signal region exceeded 6 times the noise RMS, to suppress pile-up effects (about 3\,\text{\textperthousand} of total events);   
    \item the signal corresponds either to the baseline of the non-triggering pixels, due to the logical OR, or to no-gain events due to particles impinging at the pixel borders. These two contributions together account for about 1\,\text{\textperthousand} of the total events;
    \item the signal amplitude exceeded the dynamic range of the oscilloscope to avoid saturated signals and hence a bias during the time-walk correction (about 6\% of total events).
\end{itemize}

\subsection{Results}
\label{sec:results_tb}
\begin{figure}[!ht]
    \centerline{\includegraphics[width=1\linewidth]{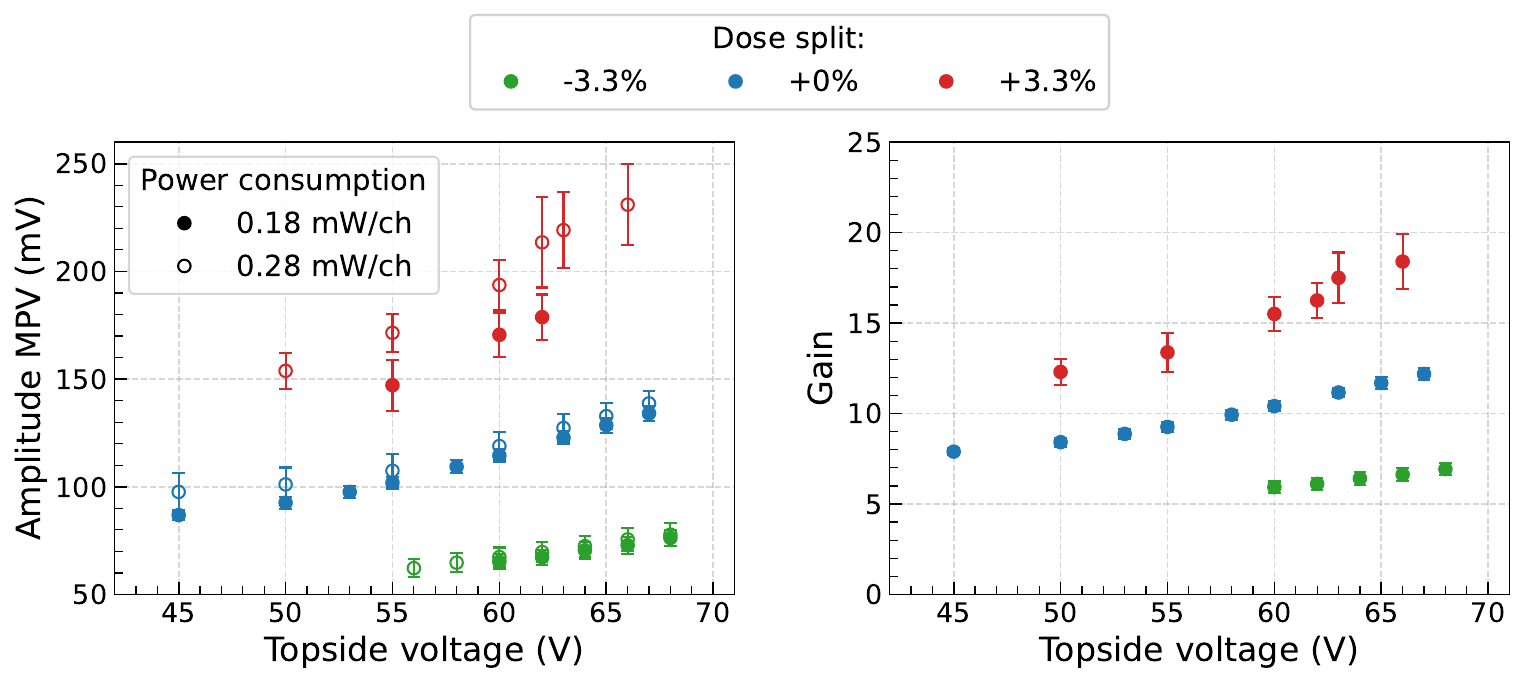}}
    \caption{(Left) MPVs of the maximum amplitude distributions measured for three different MADPix devices as a function of the topside voltage. For each device, both the standard configuration (0.18~mW/ch) and the configuration with increased supply voltage of the electronics (0.28~mW/ch) are shown. (Right) Gain as a function of the topside voltage measured for three sensors with different gain layer dose splits. \\
    InIn both the plots, the reported values correspond to the average over three pixels, while the associated uncertainty is defined as the semi-dispersion among them. Moreover, the backside electrode was biased at -25~V and the guard-ring at 9~V. }
    \label{fig:MPV_Gain}
\end{figure}

\paragraph{Amplitude and gain evaluation}\mbox{}\\
Each waveform recorded by the oscilloscope was analyzed to reconstruct the amplitude distribution and estimate the gain for different gain dose splits.
The signal amplitude is defined as the difference between the maximum value of the waveform and its baseline. The Most Probable Value (MPV) was extracted by fitting the distribution with a convolution of a Landau and a Gaussian function.
The average of the MPVs over three pixels are shown in Fig.~\ref{fig:MPV_Gain}~(left) as a function of the topside collection electrode voltage for the three different gain layer dose splits, namely: -3.3\%, +0\%, +3.3\%. A clear trend can be observed: the MPV rises as the topside bias voltage increases. This is caused by a stronger electric field within the gain layer, which reflects the corresponding enhancement in the gain. Different splits are characterized by clearly distinct amplitudes: higher gain doses correspond to larger signals. Moreover, the results were evaluated both in the standard configuration and the one with increased supply voltage of the electronics. In the latter case, higher amplitudes can be observed, because of an increase in the dynamic range of the electronics, and consequently of the front-end gain. 
The larger error bars that can be observed for the +3.3\% gain layer dose split can be justified by the results obtained with the $^{55}$Fe source, as displayed in Fig.~\ref{fig:amp_55Fe}. Indeed, the spread among the pixel signal amplitudes strongly depends on the bonding configuration, which was different for the three characterized sensors. In particular, for the sensor with +0\% and -3.3\% gain layer dose splits, the readout pixels belonged to the same column (using the coordinate-based labelling: the pixels were (4,5), (4,6), (4,7)). On the other hand, for the +3.3\% gain dose split, the acquired pixels were distributed across two different rows ((2,6), (2,7), and (3,6)), to observe the effect of the readout electronics also during the test beam measurements. As a result, they were more affected by the observed decrease in electronics power across the matrix, as shown in Fig.~\ref{fig:amp_55Fe}. An additional contribution to the larger error bars, observed in the +3.3\% gain layer dose split, arises from the gain itself. Indeed, non-uniformities in the electronic amplification are further enhanced at higher gain doses, as they scale proportionally with the input charge multiplied by the gain factor.\\
Finally, the relative spread among the three pixels observed in the test beam characterizations is consistent with the $^{55}$Fe measurements reported in Section~\ref{sec:Fe}.\\

\noindent The gain was estimated by normalizing the MPV of the amplitude to the value measured from a reference sensor without the gain layer, equal to 11~mV. Figure~\ref{fig:MPV_Gain}~(right) shows the gain values as a function of the topside voltage. As expected, the gain increases with the applied voltage. Moreover, the different gain layer dose splits exhibit well-distinguishable gain values within the expected range. For instance, at 60~V in the standard electronics configuration, the gain varies from approximately 6 to 16 depending on the gain dose, confirming the simulated values with TCAD.

\begin{figure}[ht]
    \centerline{\includegraphics[width=0.60\linewidth]{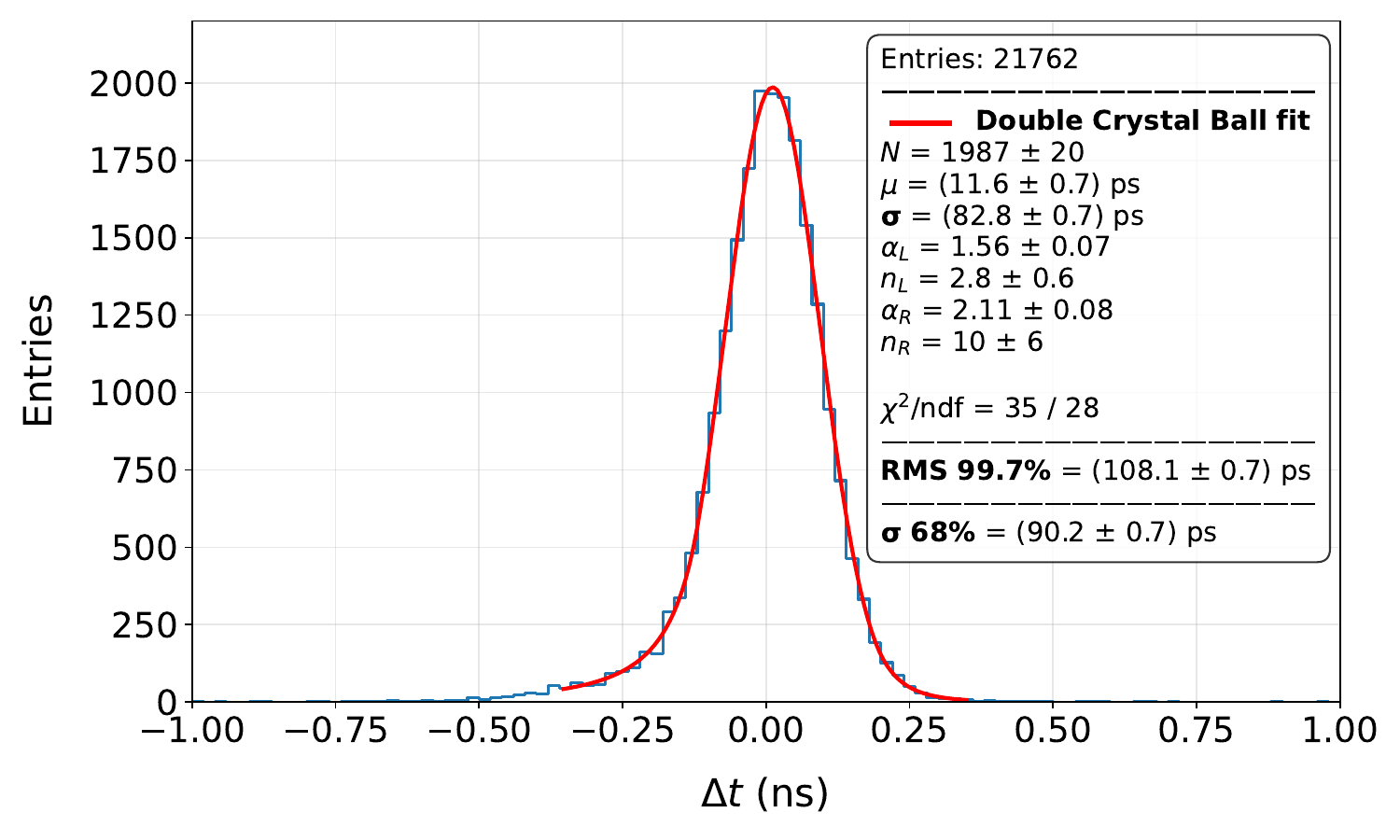}}
    \caption{Time residuals measured for one pixel in the 0.28~mW/ch front-end configuration, with a superimposed double-Crystal Ball fit. The parameter $N$ represents the normalization factor, $\mu$ corresponds to the peak position of the Gaussian core, $\sigma$ is the standard deviation of the Gaussian core, $\alpha$ determines the point of transition into a power-law behavior and $n$ controls the slope of the power-law tails on both sides. In addition to the $\sigma$ from the fit, two other numerical estimators are highlighted: the RMS evaluated on the central 99.7 percentiles of the distribution and the standard deviation $\sigma$ extracted from the central 68\% of the dataset. The topside voltage was biased at 66~V, the backside electrode at -25~V and the guard-ring at 9~V.}
    \label{fig:time_residuals}
\end{figure}

\paragraph{Time resolution}\mbox{}\\
To evaluate the timing performance of the DUT, the Time of Arrival (ToA) of each particle was extracted for both the monolithic sensor and the LGAD reference. For what concerns the LGAD plane, the ToA was determined using the crossing time at a fixed fraction of the amplitude (the Constant Fraction or CF algorithm). On the other hand, consistent with the approach followed for the laser measurements (see Section~\ref{sec:laser}), the ToA of the DUT has been evaluated using the LE algorithm, where different fixed amplitude thresholds (thr) are applied to the signals. The sensor time residuals were then computed as the difference between the timing reference and the time when the DUT signal crosses a fixed amplitude threshold:
\begin{equation} \label{eq:time_residuals}
    \Delta t = t^{\textrm{LGAD}}_{\textrm{fixed CF}} - t^{\textrm{DUT}}_{\textrm{fixed thr}} \ .
\end{equation}
As performed in Section~\ref{sec:laser}, the time-walk effect was corrected by applying an amplitude-based correction. Next, three estimators for the time residual distribution width have been computed: the RMS evaluated on the central 99.7\% of the distribution, the standard deviation $\sigma$ extracted from the central 68\% of the dataset, and the $\sigma$ obtained from a double-Crystal Ball function~\cite{double-crystalball}. Figure~\ref{fig:time_residuals} shows a time residual distribution for one pixel of the DUT with +3.3\% gain dose at a topside voltage of 66~V.

\noindent The threshold applied to extract the DUT ToA was fixed at 80~mV, as it yields the best performance.
The distribution of time residuals exhibits a Gaussian-like core with significant non-Gaussian tails on both sides, reflecting events with either a delayed or early ToA. Within the $\pm$~3~RMS fit range, the tails account for $\sim$10\% of the total distribution, with a dominant contribution from the left tail.
This combination of a sharp central peak and extended tails motivates the use of a double-Crystal Ball fitting function~\footnote{The Crystal Ball function naturally models a Gaussian core with power-law tails, and its double version allows for an independent control of each side of the tails. }.
In particular, the left-side tail is more pronounced and corresponds to the events where the ToA of the DUT is delayed with respect to the reference ones, resulting in negative time residuals. These events have been identified from both simulations and test beam measurements as originating from delayed signals at the pixel borders~\cite{Follo_2026}.\\
\noindent For a single pixel, the final time resolution ($\sigma_t$) extracted from Fig.~\ref{fig:time_residuals} after the subtraction of the LGAD contribution is: (103.6~$\pm$~1.1)~ps based on the central 99.7 percentiles of the distribution, (84.6~$\pm$~1.3)~ps from the $\sigma$ extracted over the central 68\% of the data and (77.5~$\pm$~1.3)~ps from the $\sigma$ of the double-Crystal Ball fit. The uncertainties were calculated using a standard error propagation.\\
Moreover, alternative analysis methods were employed to evaluate the DUT time resolution, including the CF algorithm and the LE algorithm with amplitude and Time-over-Threshold (ToT) time-walk corrections. All approaches yield consistent and comparable results.\\
Since the results are compatible, the LE algorithm is preferred as it is a compact solution that can be integrated in-pixel, unlike the constant fraction discriminator (CFD). Indeed, the latter usually requires extensive optimisation and larger area at the hardware level.  
Moreover, the DUT signals exhibit amplitude dependent rise times, preventing the CFD algorithm to be fully efficient and still requiring an additional correction for the maximum signal amplitude.\\

\begin{figure}[!ht]
    \centerline{\includegraphics[width=0.5\linewidth]{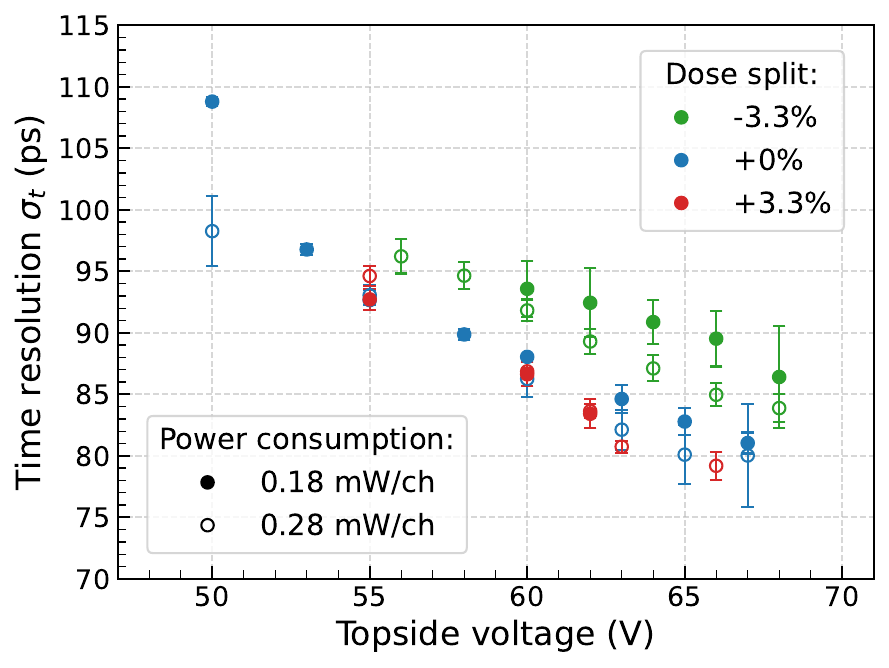}}
    \caption{Time resolution evaluated as the $\sigma$ of the double-Crystal Ball fit as a function of the topside voltage for three sensors with different gain-dose spittings. For each device, both the standard configuration (0.18~mW/ch) and the configuration with increased electronics supply voltage (0.28~mW/ch) are presented. The $\sigma_t$ values correspond to the average over three pixels, while the associated uncertainty is taken as the semidispersion among them. The backside electrode was biased at a fixed voltage of -25~V, while the guard-ring was set to 9~V. The LGAD timing contribution of about 30~ps was already subtracted for all of the data points.}
    \label{fig:time_resolution}
\end{figure}

\noindent In the following, the LE algorithm with amplitude-based correction and the double-Crystal Ball fit were chosen to extract the time resolution for all the acquired bias voltage configurations. Indeed, this function allows for the separation of the Gaussian core contribution, corresponding to the best-performing region of the sensor, from the tails, which instead account for the non-optimal signals. Figure~\ref{fig:time_resolution} reports the time resolution obtained after the subtraction of the LGAD contribution as a function of the topside voltage evaluated for the three gain layer dose splits. Each timing value corresponds to the average over three pixels.\\ 

\noindent The observed trend is consistent with expectations: as the bias voltage increases, the time resolution improves due to the establishment of a stronger electric field within the sensor, also reflecting an enhancement in the gain. Moreover, by reducing the dose split to -3.3\%, a slight degradation of the time resolution is observed. This behavior can be attributed to an increase in the jitter term, arising from the lower internal gain and the consequently reduced signal slope at the threshold crossing for a given noise level. Conversely, no significant variation is evident between the timing performance obtained with the reference doping (+0\%) and that corresponding to a +3.3\% of gain dose, indicating that, within this range, the signal-to-noise ratio contribution remains essentially unchanged.\\
Overall, the time resolution obtained in the electronic configuration with a power consumption of 0.28~mW/ch is smaller than that achieved in the standard configuration. This difference, however, progressively decreases for sensors with higher gain layer doses. The trend can be explained by the reduced impact of electronic jitter, which no longer represents the dominant contribution to the time resolution. As shown in Fig.~\ref{fig:jitter}, the jitter becomes negligible, leaving the sensor performance as the main limiting factor.\\
At the end, the best average time resolution over three pixels, obtained using a double-Crystal Ball fit, is (79.2~$\pm$~1.1)~ps for the +3.3\% of gain dose in the 0.28~mW/ch front-end electronics configuration.

\section{Conclusions} \label{Conclusions}
The optimization and experimental characterization of the first monolithic pixel sensor prototype with internal gain, implemented in the commercial 110~nm CMOS process by LFoundry, is presented in this paper. Building onto the results of the previous ER3 run, a new ARCADIA short-loop production with an increased gain has been produced.
By electrically connecting the guard-ring structure using the FIB technique, the excessive substrate current observed in the initial prototypes was effectively mitigated, resulting in a stable and low-leakage current operation. The new short-loop sensors feature an optimized gain layer profile, designed to achieve gain values in the range of 6–20, depending on the applied topside bias and gain layer dose split.\\
\noindent Laboratory measurements with a $^{55}$Fe source allowed the evaluation of the response uniformity of pixels within one chip matrix by measuring the signal amplitude induced by photons, showing a standard deviation of about 10~mV across the tested pixels, corresponding to about 12\% of the mean value, with uniform noise levels. These results provide quantitative insight to guide design optimizations and support corrections in future ERs. Moreover, an optical characterization performed with a laser setup demonstrated that the electronic jitter contribution to the time resolution could reach $\sim$25~ps for a MIP-like signal amplitude in the optimized front-end configuration. Subsequently, test beam campaigns at the CERN Proton Synchrotron with a hadron-beam momentum of 10~GeV/c further validated the performance of these devices, confirming sensor gain values in the range of 6–20, as expected. 

\noindent This CMOS MAPS technology incorporating a gain layer exhibited a time resolution, evaluated using a double-Crystal Ball fit, of (77.5~$\pm$~1.3)~ps for a single pixel and an average of (79.2~$\pm$~1.1)~ps over three pixels, confirming the feasibility of achieving a sub-100~ps timing performance at ambient temperature. Compared to the ER3, where a time resolution of about 230~ps was measured, this represents a significant improvement. The results also show the expected enhancement in time resolution with increasing applied topside voltage, as well as a dependence on the gain dose, highlighting the importance of precise control over the internal electric field.

\noindent Furthermore, these results demonstrate that monolithic sensors with an integrated gain layer developed with the 110~nm CMOS ARCADIA technology offer a promising candidate for future high-energy physics experiments and other applications requiring fast timing. Studies on timing resolution performed at different temperatures and irradiation levels, and with a tracking telescope, aimed at assessing the in-pixel time resolution and detection efficiency, have already been carried out and the analysis is ongoing. 
A new short-loop production has already been submitted using thinner substrates, with the aim of enhancing timing resolution by exploiting reduced charge collection times and faster signal rise times. In parallel, in order to mitigate the distortion term and reduce position-dependent effects, a new engineering run will be required to achieve a more uniform and stronger electric field across the active volume.

\acknowledgments
This project has received funding from the European Union's Horizon 2020 Research and Innovation programme under GA no 101004761 and from the Italian National Institute for Nuclear Physics within the CSN5 call ARCADIA.
The authors gratefully acknowledge CERN and the PS staff for continuous supports to the users. The authors would like to express their sincere gratitude to B. Pini and F. Dumitrache, responsible for the Bonding Laboratory at INFN Torino for their support and expertise. Part of this work has been carried out at Nanofacility Piemonte Istituto Nazionale di Ricerca Metrologica (INRiM), a laboratory supported by the "Compagnia di San Paolo" Foundation, and at QR Laboratories, INRiM, a micro a nanofabrication lab.

\bibliographystyle{JHEP}
\bibliography{biblio.bib}

\end{document}